\documentclass[aps,prl,reprint,twocolumn,superscriptaddress,showpacs]{revtex4-1}

\usepackage{graphicx}

\newcommand{\povo}{Dipartimento di Fisica, Universit\`{a} di Trento, 
and I.N.F.N., Gruppo di Trento, 38123 Povo (TN), Italy}

\newcommand{\rthz}{\ensuremath{/\mathrm{Hz}^{1/2}}}

\newcommand{\figr}[1]{Fig.~\ref{#1}}
\newcommand{\eqr}[1]{Eqn.~\ref{#1}}
\newcommand{\pder}[2]{\ensuremath{\frac{\partial{#1}}{\partial{#2}}}}

\begin{document}

%\preprint{}

\title{The interaction between stray electrostatic fields and a 
charged free-falling test mass}

\author{F.~Antonucci}
\affiliation{\povo}
\author{A.~Cavalleri}
\affiliation{Istituto di Fotonica e Nanotecnologie, C.N.R.- Fondazione Bruno
Kessler, 38123 Povo (TN), Italy}
\author{R.~Dolesi}
\affiliation{\povo}
\author{M.~Hueller}
\affiliation{\povo}
\author{D.~Nicolodi}
\affiliation{\povo}
\author{H.~B.~Tu}
\affiliation{\povo}
\author{S.~Vitale}
\affiliation{\povo}
\author{W.~J.~Weber}
\affiliation{\povo}

\date{\today}

\begin{abstract}
We present an experimental analysis of force noise caused by stray
electrostatic fields acting on a charged test mass inside a conducting
enclosure, a key problem for precise gravitational experiments.
Measurement of the average field that couples to test mass charge,
and its fluctuations, is performed with two independent torsion pendulum
techniques, including direct measurement of the forces caused by a
change in electrostatic charge. We analyze the problem with an improved
electrostatic model that, coupled with the experimental data, also
indicates how to correctly measure and null the stray field that
interacts with test mass charge. Our measurements allow a conservative
upper limit on acceleration noise, of 2~fm/s$^2$\rthz\ for frequencies
above 0.1~mHz, for the interaction between stray fields and charge in
the LISA gravitational wave mission.

\end{abstract}


% 91.10.Pp geodetic techniques
% 41.20.Cv electrostatics
%\keywords{}

\maketitle

Limiting stray forces on a test mass (TM) is crucial for precise
experimental gravitation, from gravitational wave (GW) observation
\cite{ugolini,GEO_charge,big_book} to tests of the equivalence principle
\cite{MICRO_ACTA_2007}, short range gravity
\cite{kapler_prl,short_long_nature}, and relativistic gyroscope
precession \cite{GPB_PRL,GPB_patch}. In all these experiments, electrostatic
force noise is cited as a precision-limiting effect. 

The orbiting GW observatory LISA (Laser Interferometry Space
Antenna \cite{big_book}) requires, along its sensitive $x$ axis,
free-fall to within 3~fm/s$^2$\rthz\ residual acceleration -- 6~fN\rthz\
force noise -- at frequencies 0.1 - 3~mHz. For LISA and its precursor
LISA Pathfinder \cite{LTP_2009,LTP_noise}, the TM is a 46~mm gold-coated
cube, inside a co-orbiting satellite and shielded, without mechanical
contact, by the gold-coated surfaces of a capacitive position
sensor \cite{sens_LISA_symp, spie_sens}, which can also apply actuation
voltages.

While the sensor is nominally an equipotential shield, two factors can
produce electrostatic forces relevant at the fN-level. First, the
floating TM accumulates charge from cosmic and solar particles, with an
expected net rate of order 50~$e$/s \cite{araujo_2005}. Second, real
metals display stray potential differences
\cite{camp_DC_japp,clive_prl} between different points on a
single conducting surface. These arise in different exposed crystalline
facets and surface contamination. Typical observed average potential
differences between roughly centimeter-size regions of a gold surface are of
order 10-100~mV \cite{pendulum_prl,cqg_pendulum,pollack_prl,norna_kelvin}. 

By itself, TM charge $q$ creates a force gradient coupling to spacecraft
motion, requiring periodic discharge \cite{diana_charge_disturb}. Stray
potentials $\delta V$ also create force gradients, whose $d^{-3}$ -- or
stronger \cite{clive_patch} -- dependence on the TM - sensor gap $d$,
motivates large,  several millimeter, gaps for LISA. 
Fluctuations in $\delta V$ also create force noise. 

\begin{figure}[t]
\includegraphics[bb=285 497 552 592,width=8cm]{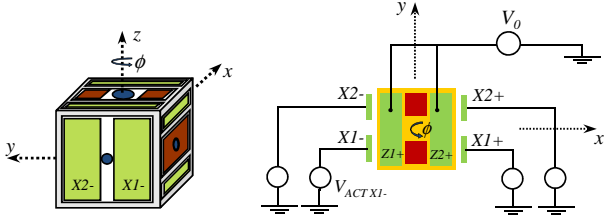}
\caption{Capacitive sensor, including (right) $X$ and $Z$ electrode
connections in the $Z$-modulation experiments. The TM - electrode gaps
are 4, 2.9, and 3.5~mm on, respectively, the $X$, $Y$, and $Z$ faces.
The presented torsion pendulum measurements detect the rotation $\phi$.}
% NB generated with MS Publisher, eps, high quality figure
\label{sensor_fig}
\end{figure}

The mutual interaction between charge and stray potentials 
\cite{shoe_noise,tuck_budget,pete_lowf,DC_bias_asr} can
be written, to linear order in $q$, 
\begin{equation}
F_x 
=
- \frac{q}{C_T} \left| \pder{C_X}{x} \right| \Delta_x \, \, .
\label{force_def}
\end{equation}
$\Delta_x$ is an effective potential difference proportional to
$\pder{F_x}{q}$ and will be calculated shortly. $C_X$ and
$C_T$ are, respectively, the TM capacitances to an $X$-electrode and the
entire sensor (see \figr{sensor_fig}).

This interaction produces force noise in two ways.  First, any residual
$\Delta_x$ multiplies random charge noise,
\begin{eqnarray}
S_{F(\delta q)}^{1/2} 
& =  & \frac{S_q^{1/2}}{C_T} \left| \pder{C_X}{x} \right| \Delta_x 
% \nonumber 
\\
& \approx & 
\! \! 7 \, \mathrm{fN/Hz^{1/2}}   
% \nonumber \\ & \: \: \: \times &
\!  \times \! 
\left( \! \frac{\Delta_x }{0.1 \: \mathrm{V}} \! \right) \! \!
\! \left( \! \frac{\lambda_{eff}}{300 \: \mathrm{/s}}  \! \right)^{1/2} \! \!
\! \left( \! \frac{10^{-4} \: \mathrm{Hz}}{f} \! \right)
.
\nonumber
\label{force_noise_q}
\end{eqnarray}
$\Delta_x \approx 100$~mV is typical for LISA 
prototype sensors 
\cite{pendulum_prl,cqg_pendulum,cqg_4TM_EM}. $\lambda_{eff}$ is the
equivalent single charge event rate that gives a ``red'' Poissonian
shot noise $S_q = \frac{2e^2 \lambda_{eff}}{\omega^2}$, estimated 
at roughly 300~/s \cite{araujo_2005}, larger during solar
flare events \cite{helios_2004}. This random charge force noise 
$S_{F(\delta q)}$ can be eliminated by
nulling $\Delta_x$ with applied voltages \cite{DC_bias_asr,pendulum_prl}.

Second, fluctuations in $\Delta_x$ will multiply any nonzero TM charge to 
produce force noise, 
\begin{eqnarray}
S_{F(\delta\Delta_x)}^{1/2} 
 & = & \frac{q}{C_T} \left| \pder{C_X}{x} \right| S_{\Delta_x}^{1/2}  \\
 & \approx & 1.3 \: \mathrm{fN/Hz^{1/2}} \times
\left( \frac{q}{10^7 \: e} \right) \!
\left( \frac{S_{\Delta_x}^{1/2}}{ 100 \: \mathrm{\mu V/Hz^{1/2}} } \right)
.
\nonumber
\label{force_noise_Delta_x}
\end{eqnarray} 
10$^7$ $e$ is roughly two days of accumulated charge and 
a reasonable discharge threshold.  

This Letter addresses these two sides of the $q$-$\delta V$
interaction. Our analysis, considering spatial surface potential
variations on both the TM and sensor, highlights systematic errors in
measuring $\Delta_x$ with applied electrostatic fields, consistent with
our experimental data. Force noise from field fluctuations is then
addressed by measurements of stray potential fluctuations. Experiments
employ a hollow LISA-like TM suspended as a torsion pendulum inside a
prototype LISA capacitive sensor connected to a prototype sensing and
actuation electronics \cite{pendulum_prl}. All relevant surfaces have been
sputtered with gold and held under vacuum for more than a year. 
Our measurements
in this flight-realistic configuration allow a conservative upper
limit for the TM acceleration noise caused by the interaction between charge and
stray fields. 

\begin{figure}[b]
\includegraphics[bb=294 481 544 606,width=8cm]{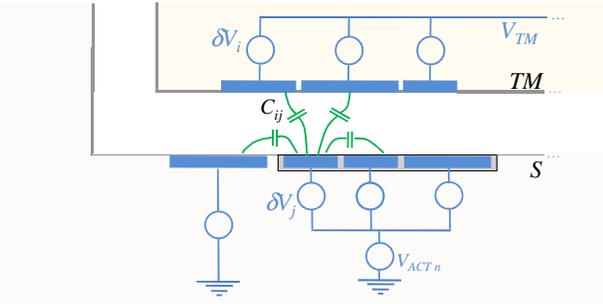}
\caption{
Schematic of the electrostatic model, with capacitively coupled sensor (S) and 
TM domains.}
% NB generated with MS Publisher, eps, high quality figure
\label{cap_model_fig}
\end{figure}

The electrostatic interaction is modeled as a patchwork 
of discrete TM and sensor (S) surface
domains at potentials $V_i$ (see \figr{cap_model_fig}), coupled
by capacitors $C_{ij}$, with
$q_i = \sum_j{C_{ij}\left( V_i - V_j \right)}$ 
\footnote{This is equivalent to the
capacitive matrix formulation, with $q_i = \sum_j{c_{ij} V_j}$ and 
energy $U = \frac{1}{2}\sum_{i,j}{c_{ij} V_i
V_j}$. The symmetries imposed by the
Laplace equation \cite{cap_herrera} allow the electrical circuit
analogy, with $c_{ij} = c_{ji} = - C_{ij}$ for $j \neq i$
and $c_{ii} = \sum_{j\neq i}{C_{ij}}$. From this \eqr{estat_force} follows.}. Stray
potentials are defined by ideal generators $\delta V_i$.
For a sensor domain, $V_i = \delta V_i$ or, if
located on an electrode attached to generator
$V_{ACTn}$, 
$V_i = \delta V_i + V_{ACTn}$. For a TM domain, $V_i = \delta V_i +
V_{TM}$, where $V_{TM}$ is an effective average TM 
potential \footnote{$We choose the convention 
$\displaystyle \sum_{i(TM),j(S)}{C_{ij} \delta V_i} = 0$, 
such that $V_{TM}$ = 0 when
$q = 0$ and all sensor domains are grounded, $V_{Sj} =
0$.},
\begin{equation}
V_{TM} = \frac{q}{C_{T}} + \frac{\sum_{j(S)}{C_{Sj} V_{j}}}{C_{T}} ,
\label{estat_vm}
\end{equation}
where index $j(S)$ restricts the sum to sensor domains.  
$C_{Sj} \equiv \sum_{i(TM)}{C_{ij}}$ is the total capacitance
between sensor domain $j$ and all TM domains, and $C_{T} \equiv
\sum_{j(S)}{C_{Sj}}$ is the total TM capacitance to the sensor.  

The force on the TM along the $x$ axis is
\begin{equation}
 F_x = \frac{1}{2} 
\sum_{i,j<i}{\pder{C_{ij}}{x} \left( V_i - V_{j} \right)^2 } \, ,
\label{estat_force}
\end{equation}
summing over all domain pair capacitances, including those with
nearby domains on both the TM or sensor \cite{nico_lisa_symp}. This
combines with \eqr{estat_vm} to yield the charge dependent
force.  For a centered TM, such that $\pder{C_T}{x} = 0$,
\begin{equation}
\pder{F_x}{q}
\equiv
\frac{-1}{C_T}
\left| \pder{C_X}{x} \right| \Delta_x 
=
\frac{-1}{C_T} \! \! \!
\sum_{i \left(TM \right),j \left( S \right) }
{\! \! \! \pder{C_{ij}}{x} \left( V_j  - \delta V_i \right) }
.
\label{force_q_Delta_x}
\end{equation}
The derivative $\pder{C_X}{x}$ normalizes $\Delta_x$ to a single $X$
electrode potential, such that $+V$ applied to electrode $X1+$ or $X2+$
increases $\Delta_x$ by $V$, with the opposite change obtained with
$X1-$ or $X2-$. 
Equation \ref{force_q_Delta_x} differs from the analogous
formula in Ref. \cite{DC_bias_asr} as it includes the spatially varying
TM potential. 

Shear forces arise naturally in \eqr{estat_force}, in
contrast with equipotential TM models\cite{diana_charge_disturb, DC_bias_asr},
for which the relevant derivative in \eqr{estat_force} becomes
$\pder{C_{Sj}}{x}$. Nonzero for the gap-dependent capacitances of a
sensor $X$-face domain, $\pder{C_{Sj}}{x}$ vanishes for a
typical sensor $Y$ or $Z$ domain far from the TM edge, as TM motion
along $x$ gives fixed-gap sliding of a large conducting plane. With a
patchwork TM surface, a sensor $Y$ or $Z$ domain ``overlaps'' with
several opposing TM domains, giving $\pder{C_{ij}}{x} \neq 0$ and thus
a force in the $x$ direction. The field component along the underlying conducting TM
surface vanishes, but its gradient does not, creating a shear force on
the TM surface dipole distribution that generates the varying surface
potential \cite{clive_patch,jackson}. Stray torques from such shear
forces limited sensitivity for the spherical Gravity Probe B gyroscopes
\cite{GPB_PRL}.

Electrostatic shear is not essential to $\pder{F_x}{q}$ (see Ref.
\cite{supp}); $\Delta_x$ reflects the average field along $x$ felt by
the TM free charge, and the uniform change in $V_{TM}$ caused by $q$ does not
create significant field gradients that shear the TM surface dipoles.
However, shear forces impact attempts to measure $\Delta_x$ with applied
voltages, and thus also the random charge problem. 

Ideally, $\Delta_x$ is measured by the force caused by a change
in TM charge, with voltages then applied to the $X$
electrodes to null $\pder{F_x}{q}$. An easier proposed 
method \cite{DC_bias_asr}
simulates charge by modulating $V_{TM}$ with $V_{0}
\sin {2 \pi f_0 t}$ applied to the 4 $Z$ electrodes. Combining Eqns.
\ref{estat_vm} and \ref{estat_force} yields the coherent force 
\begin{eqnarray}
 F_{x(1f)} = - V_{0} \sin{2 \pi f_{0} t} 
\times
\nonumber \\
\left\{ 
\alpha_z \left| \pder{C_X}{x} \right| \Delta_x 
- 
\sum_{i(TM),j(S_z)}{ \pder{C_{ij}}{x} \left( \delta V_{j} - \delta V_{i} \right) }
\right\} 
.
\label{measure_eqn}
\end{eqnarray}
Here, $j\left(S_z \right)$ sums over domains on the modulated $Z$
electrodes and $\alpha_z = 4 {C_Z} / C_T \approx 0.07$. 
The first term is proportional to $\Delta_x$, while
the second, irrelevant to \pder{F_x}{q}, is the shear action of
modulated field gradients near the $Z$ electrodes on nearby TM surface
dipoles. It vanishes with an equipotential TM, with $\pder{C_{Sj}}{x}
\approx 0$ for a Z electrode domain $j$.

TM inclination with respect to the $Z$ electrodes also introduces error,
with a gap-varying \pder{C_Z}{x}\ coupling the $Z$-electrode surface
potentials and modulation voltage. However, shear coupling to the
varying TM potential represents a more fundamental error that limits any
technique to measure \pder{F_x}{q} with applied voltages
-- without actually varying $q$ -- even with perfect alignment.

Experimentally, with a torsion pendulum sensitive to torque, $N_{\phi}$
(see \figr{sensor_fig}), we assess $\Delta_x$ and its fluctuations by
measuring the rotational imbalance $\Delta_{\phi}$ relevant to
$\pder{N_{\phi}}{q}$. $\Delta_{\phi}$ is defined analogously to
$\Delta_x$ ($x \rightarrow \phi$ in Eqns. \ref{force_def},
\ref{force_q_Delta_x}, and \ref{measure_eqn}). With an equipotential TM
and individually equipotential electrodes, $\Delta_x$ and
$\Delta_{\phi}$ become, respectively, the left-right and diagonal
imbalances of the same 4 $X$-electrode potentials, $\Delta_x = \left(
V_{X1+} + V_{X2+} - V_{X1-} - V_{X2-} \right)$ and $\Delta_{\phi} =
\left( V_{X1+} - V_{X2+} - V_{X1-} + V_{X2-} \right)$
\cite{DC_bias_asr}. With electrostatically inhomogeneous conductors,
$\Delta_{\phi}$ has gap-varying sensitivity to the $Y$ surface
potentials as well as the $X$ domains that dominate $\Delta_x$
\cite{supp}, and thus statistically overestimates $\Delta_x$ and its
fluctuations.

\begin{figure}
% go from 0.65 to 0.7 for scale size
\includegraphics[scale=0.7]{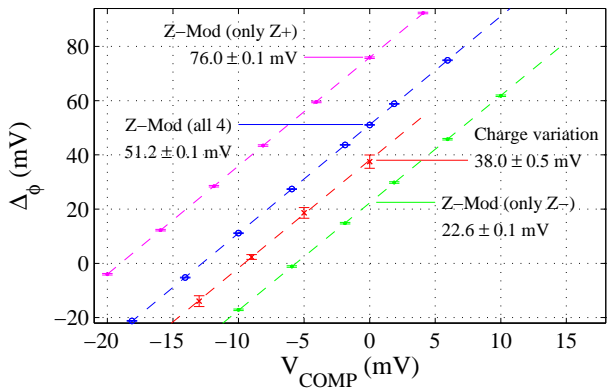}
\caption{
Comparison of $\Delta_{\phi}$ obtained by charge
variation and by the $Z$-modulation technique, 
with extracted values for the uncompensated 
$\Delta_{\phi}$ ($V_{COMP} = 0$).  Single point 
error bars ($<$~300~$\mu$V) are not visible in the 
modulation data.  }
\label{charge_scan_fig}
\end{figure}

The Z-modulation measurement of $\Delta_{\phi}$ is compared here 
with direct measurement of $\pder{N_{\phi}}{q}$ 
(see \figr{charge_scan_fig}).
Measurements are performed as a function of compensation voltage
$V_{COMP}$, applied with positive (negative) polarity on the $X1+$ and $X2-$
($X1-$ and $X2+$) electrodes. We measure $\pder{N_{\phi}}{q}$ by the change in
torque, up to 5~fN\,m and measured to $\pm$0.1~fN\,m, upon rapid change
in TM charge, of order 10$^7 e$, caused by 10-30~s UV illuminations. The
$Z$-modulation and charge measurement technique were applied and
analyzed as in Refs. \cite{DC_bias_asr,pendulum_prl}. 

The uncompensated $\Delta_{\phi} \approx 38$~mV measured by charge
variation is in the typical range of other
sensors \cite{pendulum_prl,cqg_pendulum}, and the slope of 4 confirms
Eqns. \ref{force_q_Delta_x}-\ref{measure_eqn} for 4 compensated $X$
electrodes. The result for 4-electrode $Z$-modulation, $\Delta_{\phi}
\approx 51$~mV, is 13~mV larger. Additional tests with modulation on
only the $Z+$ or $Z-$ electrode pairs give values of $\Delta_\phi$
differing by more than 50~mV (\figr{charge_scan_fig}), with a 90~mV
range observed for $\Delta_\phi$ with individual modulation of the 4 $Z$
electrodes. 

% \footnote{Capacitive parameters \pder{C_X}{\phi} and
% $C_T$, evaluated by FEM \cite{nico_lisa_symp}, affect both the
% calculated TM charge and $\Delta_{\phi}$ in such a way as to not affect
% the final slope determination.}

The disagreement of the various $Z$-modulation results with
$\Delta_{\phi}$ measured with $\pder{N_{\phi}}{q}$ indicates the level of
error in the $Z$-modulation technique.
With an equipotential TM, these measurements should yield the same
value. The variation between results with different $Z$ electrodes
reflects differences in the TM potentials near the different
$Z$-electrodes. This partially averages out by modulating all 4
$Z$-electrodes, but still leaves a 13~mV deviation from the true
$\Delta_{\phi}$. For comparison and an indication of long term
stability, the same sensor 1 year before gave $\Delta_{\phi} \approx
135$~mV, with an 8~mV difference between the \pder{N_{\phi}}{q} and
$Z$-modulation techniques.

\begin{figure}[b]
\includegraphics[scale=0.65]{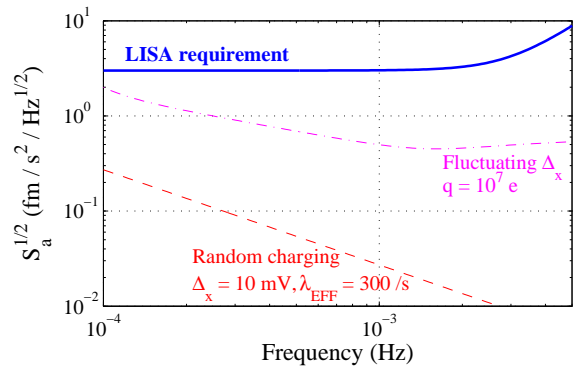}
\caption{
Conservative acceleration noise budget for the $q$-$\delta V$
interaction, with (bold) the LISA goal.}
\label{es_budget}
\end{figure}

To limit the random charge contribution to the LISA acceleration noise
(\figr{es_budget}), we want $\Delta_x < 10$~mV. This will likely
require in-flight measurement and 
compensation of intrinsic imbalances typically of order 100~mV, 
repeated periodically, given the slow drifts observed here and
elsewhere \cite{pollack_prl}. Additionally, analysis and measurements
indicate that errors associated with the $Z$-modulation technique may
not allow 10~mV accuracy. This would require the more cumbersome direct
measurement of \pder{F_x}{q} in flight, which needs UV light actuation,
charge measurements, and transient force detection. 

Noise in $\Delta_x$ is assessed with two different measurements of
$S_{\Delta_{\phi}}$. We first measure torque noise with a charged TM,
attributing any excess to $\Delta_{\phi}$ fluctuations,
\begin{equation}
S_N \left( q \right) - S_N \left( 0 \right) 
= \left[ \frac{q}{C_T} \left| \pder{C_X}{\phi} \right| \right]^2 
S_{\Delta_{\phi}}
\: \: \: .
\label{Delta_phi_noise}
\end{equation}
Measurements for 3 consecutive weekends with the TM charged to $V_{TM} =
1.82 \pm 0.02$~V ($q \approx 4 \times 10^8 \,e$) were sandwiched between 4 weekends
with the TM neutral to within 20~mV. The noise analysis, similar to Ref.
\cite{cqg_fused}, uses 25000~s Blackman-Harris windows with 66\% overlap
-- 55 and 70 windows for, respectively, the charged and neutral TM data
-- binned into 8 frequencies per decade and averaged, with uncertainties
based on standard deviation among pre-averaged groups of 5 windows. 

\begin{figure}
% go from 0.65 to 0.7 scale
\includegraphics[scale=0.65]{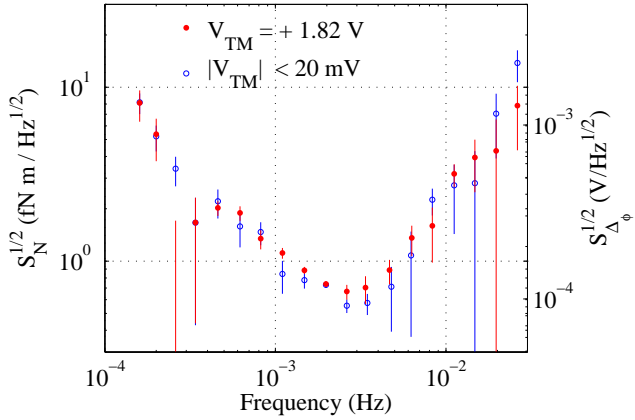}
\caption{
Pendulum torque noise measured with the TM nearly neutral
and when charged to a potential $V_{TM}$ = +1.82 V, with conversion 
into $S_{\Delta_{\phi}}$ shown at right.}
\label{S_VTM_fig}
\end{figure}

\begin{figure}
% go from 0.65 to 0.7
\includegraphics[scale=0.65]{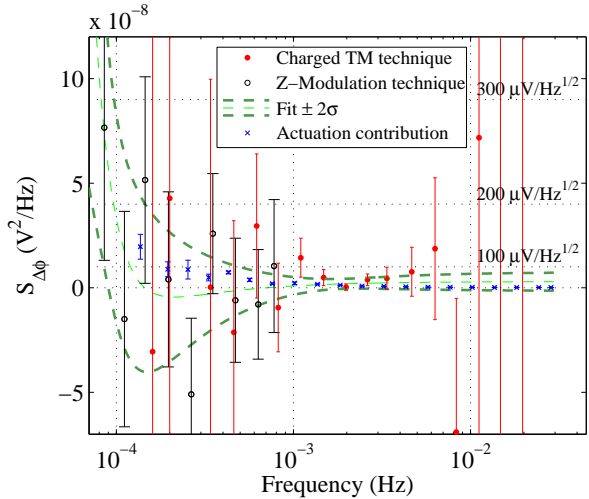}
\caption{Stray potential noise $S_{\Delta_{\phi}}$,
measured with two techniques.  Statistically
insignificant large error points are omitted from each dataset.  Also
shown are a fit to the joint dataset and the noise contribution from the 
actuation electronics.}
\label{S_Delta_phi_fig}
\end{figure}

Figure \ref{S_VTM_fig} shows averaged torque noise, similar for the
charged and neutral TM and with a minimum near 3~mHz of roughly
0.7~fNm\rthz (120~$\mu$V\rthz). Following \eqr{Delta_phi_noise}, we
subtract the neutral TM background $S_N \left( 0 \right)$ -- measured to
be stationary at the 0.1 (fN m)$^2$/Hz level -- to obtain
$S_{\Delta_{\phi}}$ in \figr{S_Delta_phi_fig}. The 1-4~mHz average is
roughly 50~$\mu$V\rthz, resolved at nearly the 2$\sigma$ level.
Background noise, and the associated errors bars, increase at both
higher and lower frequencies, with no resolvable excess. 

We also measure residual fluctuations in the $Z$-modulation signal, as
in Ref. \cite{cqg_pendulum} and \eqr{measure_eqn}, with $\Delta_{\phi}
\left( t \right)$ detected in the coherent torque amplitude at the
modulation frequency $f_0$. This overestimates $S_{\Delta_{\phi}}$ by
the second term in \eqr{measure_eqn}, but improves sensitivity to low
frequency fluctuations in $\Delta_{\phi}$, which are spectrally
upshifted to amplitude modulate the torque carrier around $f_0$ = 3~mHz,
chosen to minimize torque noise. $V_{COMP}$ is adjusted to null the
signal upon starting the measurement. $\Delta_{\phi} \left( t \right)$
is corrected for the measured dependence on tilt-induced TM translation
inside the sensor. We subtract background measurement noise, typically
500~$\mu$V\rthz\, as calculated with the demodulated quadrature (cosine)
torque phase, which contains statistical torque noise without
electrostatic signal. Spectra are calculated with 60000~s windows,
binned and averaged, and then background subtracted for each of 13
weekend measurements, with uncertainties based on scatter between
different windows. Figure \ref{S_Delta_phi_fig} shows a weighted mean of
these data.

The measurements with a charged TM ($\bullet$) and Z-modulation
($\circ$) combine in \figr{S_Delta_phi_fig} \footnote{For both measurements, 
the estimated noise power excess
$S_{\Delta_{\phi}}$ is in many cases smaller than the measurement
uncertainty, and so negative data points are statistically inevitable.
These negative data do not allow a conventional log-scale plot of linear
spectral density $S_{\Delta_{\phi}}^{1/2}$ and force a linear plot of
the power spectral density. Key linear spectral levels like 100
$\mu$V/Hz$^{1/2}$ are shown as a guide. } for a significant upper limit on the noise
power $S_{\Delta_{\phi}}$ from 0.1 to 5~mHz. We fit the combined dataset
to various models with a low frequency increase -- Fig.
\ref{S_Delta_phi_fig} shows a fit with $f^{-1}$ and $f^{-4}$ terms --
and find 2$\sigma$ confidence intervals of $\left( 0 , 80 \right)
\mu$V\rthz\ at 1~mHz and $\left( 0, 290 \right) \mu$V\rthz\ at 0.1~mHz.
Also shown ($\times$) is the contribution from the 4 $X$-electrode
actuation circuits \footnote{The actuation noise, measured with a
chopper-demodulation technique, comes from commercial 
digital-to-analog converter (NI-6703) and
instrumentation amplifier (AD-622) components.}, measured separately to
give $f^{-1}$ noise power with roughly 50~$\mu$V\rthz\ at 1~mHz. The
weakly detected fluctuations are thus
consistent with electronic noise, and we do not resolve true surface
potential fluctuations in the LISA bandwidth. $\Delta_{\phi}$
effectively sums the noise of many ($\approx$~16, see Ref. \cite{supp})
areas the size of a LISA $X$-electrode ($\approx$~500~mm$^2$), and, for
fluctuations that are uncorrelated on larger spatial scales, our
2$\sigma$ upper limit corresponds to 12~$\mu$V\rthz\ in the average
potential difference between 500~mm$^2$ surface regions. 

The +2$\sigma$ curve in \figr{S_Delta_phi_fig} is taken as a upper limit
for $S_{\Delta_x}$ in the acceleration noise budget in \figr{es_budget}.
Assuming a TM charge of 10$^7 e$, this result is compatible with the
LISA goals, marginally so at 0.1~mHz. This significantly improves upon
previous upper limits with LISA prototype hardware, from roughly
1~mV\rthz \cite{cqg_pendulum} to 80~$\mu$V\rthz\ at 1~mHz. 
The limit is also below the 150-200~$\mu$V\rthz\ deduced 
for $S_{\Delta_x}^{1/2}$ at 1~mHz from observations of 
potential fluctuations between opposing gold-coated plates 
\cite{pollack_prl,supp}.  
 
Several design aspects merit consideration for improving upon the noise
budget in \figr{es_budget}. A thin ground wire can eliminate 
TM charge \cite{touboul_nanog,MICRO_ACTA_2007}, but introduces thermal
mechanical noise well beyond the fm/s$^2$/Hz$^{1/2}$ level
\cite{willemot_rsi}. With a floating TM, the $q$-$\delta V$ interaction
has roughly $d^{-1}$ dependence (\eqr{force_def}). 
Larger gaps help, but even a factor 10, 
$d$~=~4~cm, will lower the random charge noise in \figr{es_budget} only if
$\Delta_x < 100$~mV, which may still require voltage-controlled
electrodes for compensation. This introduces actuation circuitry noise,
which sums with, or even dominates over, surface potential
fluctuations. Preliminary measurements for the LISA Pathfinder
electronics indicate a circuit contribution $S_{\Delta_x}^{1/2}
\approx$ 30~$\mu$V/Hz$^{1/2} $\cite{hev_tech_note}. 

Along with the interaction with TM charge, stray potential
fluctuations can also create force noise by mixing with stable domain
potentials, even with $q = 0$, as allowed by \eqr{estat_force}. Analysis
of this effect demands combining averaged potential fluctuation data
with the domain spatial distribution and correlations. This will impact
in-flight operation issues such as the extent to which continuous TM
discharge can reduce noise. 

\begin{acknowledgments}
This work was supported by the 
Istituto Nazionale di Fisica Nucleare, 
the Agenzia Spaziale Italiana (LISA Pathfinder contract), and
the Italian Ministry of University and Research (PRIN 2008).
\end{acknowledgments}

\appendix

\section{Supplemental material for 
{\it The interaction between stray electrostatic fields and a 
charged free-falling test mass}: Statistical comparison of 
different stray potential measurements and their application to 
the interaction with TM charge}

This supplementary report addresses how -- and which -- stray
electrostatic potentials create a force on a charged test mass (TM),
\begin{equation}
\pder{F_x}{q} \equiv -\frac{1}{C_T}\pder{C_X}{x}\Delta_x  
\nonumber
\end{equation} 
and how to 
compare different stray potential measurements 
with $\Delta_x$, and its fluctuations, for the LISA geometry. All
prototype sensors or sample surfaces will be different, even
with nominally identical gold surfaces prepared in the same way. As
such, we seek a statistical comparison between different
measurements under various assumptions for the underlying distribution
of stray potentials. The figure of merit that we will use is the mean
square variance $\langle \Delta_x^2 \rangle$ of the potential difference
relevant to charging in LISA. The noise power spectral density should
scale in the same fashion, as it is the Fourier transform of the related
correlation, $\langle \Delta_x \left( t \right) \, \Delta_x \left( t +
\tau \right)\rangle$. 

The principle conclusions of this report are:

\begin{itemize}
\item The coupling to charge $\pder{F_x}{q}$ is essentially an average 
electrostatic field along $x$.  It (and equivalently $\Delta_x$) is 
thus dominated by the stray potentials on the surfaces of the 
TM and surrounding enclosure that are normal to the $x$ axis. 

\item The rotational stray potential imbalance $\Delta_{\phi}$, 
studied experimentally in the main article, is statistically noisier
than $\Delta_x$ for stray potential distributions that 
are dominated by domains the size of the LISA electrodes (500 mm$^2$)
or smaller.

\item Average potential difference measurements performed in the
geometry of Ref. \cite{pollack_prl} are readily applicable to the LISA
geometry. To compare with $\Delta_x$, which is normalized to the size of
a single LISA $X$ electrode, those results (for linear spectral noise
density) must be multiplied by a factor 5-6 for any characteristic
domain sizes up to the dimensions ($\approx$ 2000~mm$^2$) of the
surfaces measured. 

\end{itemize}

\begin{figure}[b]
\includegraphics[bb=316 486 535 608, width=7cm]{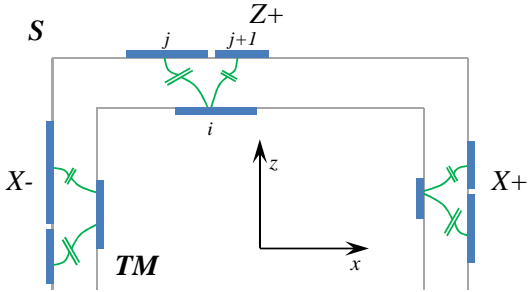}
\caption{Cartoon illustrating the capacitive coupling between 
domains on the TM
and surrounding enclosure (or sensor, $S$). Adjacent elements on the $X$
sensor faces contribute $\pder{C_{m,n}}{x}$ with the same sign, positive
on the $X+$ face and negative on $X-$, as they undergo the same gap
change upon TM motion along $x$. The $x$ dependence of the adjacent
domains on a $Z$ face cancel out to first approximation, as the decrease
in area overlap between TM domain $i$ and sensor domain $j$ is matched
by the decrease in the $\left( i , j+1 \right)$ overlap, resulting in
$\pder{C_{TMi}}{x} \approx 0$. Ultimately, this gives a weak
dependence of the charge coupling $\pder{F_x}{q}$ (or $\Delta_x$) on the
$Y$ and $Z$-face surface potentials.}
\label{estat_model_supp}
\end{figure}

In the capacitive model, the potential difference $\Delta_x$, 
\begin{equation}
\Delta_x =  
\frac{1}{\left| \pder{C_X}{x} \right|}
\sum_{i \left(TM \right),j \left( S \right) }
{\! \! \! \pder{C_{ij}}{x} \left( V_j  - \delta V_i \right) }
\label{s_Delta_x_I}
\end{equation}
can also be expressed
\begin{equation}
\Delta_x 
 =  
\frac{1}{\left| \pder{C_X}{x} \right|}
\left[
\sum_{j \left( S \right) }
{\! \! \! \pder{C_{Sj}}{x} V_j } 
- 
\sum_{i \left( TM \right) }
{\! \! \! \pder{C_{TM\,i}}{x} \delta V_i } 
\right]
\label{s_Delta_x_II}
\end{equation}
where $C_{Sj} \equiv \displaystyle \sum_{i(TM)} C_{ij}$ is the total capacitance
between sensor domain $j$ and the entire TM, and $C_{TMi} \equiv
\displaystyle \sum_{j(S)} C_{ij}$ is the total capacitance between TM domain $i$ and
the surrounding sensor. $\pder{C_{Sj}}{x}$ and $\pder{C_{TMi}}{x}$ are
positive for domains on the sensor or TM $X+$ faces and negative for
domains on the $X-$ faces, as the relevant gaps change with TM motion
along $x$. The same derivatives are nearly zero for domains on the $Y$
or $Z$ faces; TM slide motion along $x$ changes the individual
inter-domain capacitances by altering their effective overlap (see
\figr{estat_model_supp}), but the total capacitance of a domain to the
opposing surface is unchanged. This is no longer true near the TM edge,
but border effects play a secondary role for an enclosure with TM size
much larger than the relevant TM - sensor gaps, as is the case for LISA,
 with cube sidelength $s$= 46~mm and gap $d$ = 4~mm. 
Aside from the normalization factor
$\pder{C_X}{x}$, the two terms in \eqr{s_Delta_x_II} represent weighted
averages of the domain potentials on the sensor and TM $X$ faces, with
the weights provided by the capacitance derivatives, which are
proportional to area for uniform $d$. These average potentials 
determine the average residual electrostatic field along $x$, 
and thus also $\Delta_x$.

\begin{figure}[t]
\includegraphics[bb=295 463 547 626, width=7cm]{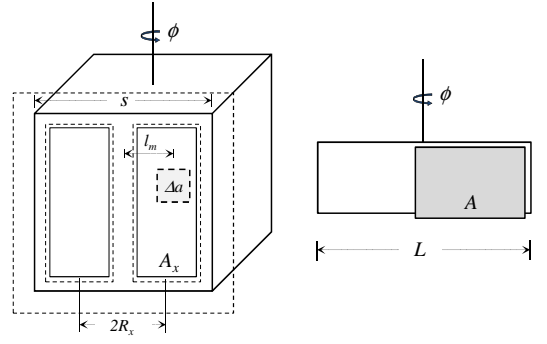}
\caption{Schematic illustration of two experimental configurations for
torque measurements of surface potential differences. At left is a LISA
prototype TM (cube sidelength $s$ = 46~mm), with the outlines of the
sensor $X$ electrode footprints, with surface area $A_x$ = 529~mm$^2$
and a on-center semi-separation $R_x$ = 10.75~mm and (dashed) guard ring
surfaces. The gap $d_x$ from TM to the $X$ electrodes and adjacent
grounded sensor guard ring surfaces is 4~mm. At right is the geometry of
Ref. \cite{pollack_prl}, in which an average potential difference is measured
between one half of the suspended plate and a facing electrode plate of
width $\frac{L}{2}$ = 57.2~mm and area $A$~=~2180~mm$^2$.}
% NB generated with MS Publisher, eps, high quality figure
\label{supp_sensor_fig}
\end{figure}

With $\Delta_x$ dominated by the average potential differences between
opposing TM and sensor surfaces, we construct an approximate statistical
model, which considers the average difference $\delta V_m$ between
opposing TM and sensor surface elements of area $\Delta a$ 
(see \figr{supp_sensor_fig}). Such
elements are not necessarily equipotential but are, first, large enough
($\Delta a > d^2$) to allow, for ease of calculation, an infinite
parallel plate model for the capacitance derivatives, $\pder{C_m}{x}
\approx \frac{\Delta a \epsilon_0}{d^2}$, such that 
\begin{equation}
\Delta_x  \approx  \frac{d^2}{A_x \epsilon_0} 
\sum_{m} \frac{\Delta a \epsilon_0}{d^2}{\delta V_m}
 = \frac{\Delta a}{A_x} 
\sum_{m} {\delta V_m}
\label{s_ave_Delta_x}
\end{equation}
where $\pder{C_X}{x} \approx \frac{A_x \epsilon_0}{d^2}$. 
Second, $\Delta a$ is considered large enough such that the average
potential is uncorrelated between elements, with $\langle \delta V_m \,
\delta V_n \rangle = \delta_{m,n} V_0^2$, where $V_0^2$ is the mean
square value of the potential difference 
averaged over surface $\Delta a$. As such,
we can estimate the statistical variance $\langle \Delta_x^2 \rangle$ 
% \begin{eqnarray}
\begin{equation}
\langle \Delta_x ^2 \rangle 
 \approx  \frac{\Delta a^2}{A_x^2} \sum_{m,n} 
\langle \delta \! V_m \, \delta \! V_n \rangle 
=  \frac{\Delta a^2}{A_x^2} V_0^2 N 
=  \Delta a \, V_0^2 \frac{2s^2}{A_x^2} 
\label{s_ave_Delta_x_2}
\end{equation}
%\end{eqnarray}
with $N \approx \frac{2 s^2}{\Delta a}$ the number of
elements on the two $X$ faces.  We will confront this variance
with other potential difference measurements under different assumptions
of the minimum area $\Delta a$ for which surface elements can be 
considered uncorrelated.  

The same arguments apply for the 
rotational imbalance $\Delta_{\phi}$, defined 
\begin{equation}
\Delta_{\phi} 
 =  
\frac{1}{\left| \pder{C_X}{\phi} \right|}
\sum_{i \left(TM \right),j \left( S \right) }
{\! \! \! \pder{C_{ij}}{\phi} \left( V_j  - \delta V_i \right) }
\label{s_Delta_phi}
\end{equation}
We can perform a similar analysis, using an infinite wedge approximation
to estimate the gap-dependent capacitance derivatives, with
$\pder{C_m}{\phi} \approx \frac{\Delta a\epsilon_0}{d^2} l_m$ for an
element of area $\Delta a$ at a distance $l_m$ from the center of the TM
face (see \figr{supp_sensor_fig}). Likewise, $\Delta_{\phi}$ normalizes
to the $X$-electrode derivative $\pder{C_X}{\phi} \approx
\frac{A_x \epsilon_0 R_x}{d^2}$. Both $X$ and $Y$ faces give
gap-dependent contributions to $\Delta_{\phi}$,
\begin{eqnarray}
\Delta_{\phi}  & \approx &  
\frac{d^2}{A_x R_x \epsilon_0}
\sum_{m} \frac{\Delta a \, l_m \epsilon_0}{d^2}{\delta V_m}
= \frac{\Delta a}{A_x R_x }
\sum_{m} {l_m \delta V_m}
\nonumber \\
\langle \Delta_{\phi} ^2 \rangle 
& \approx & 
\Delta a \, V_0^2 \frac{4s^2 \bar{l^2}}{A_x^2 R_x^2} 
\label{s_Delta_phi_model}
\end{eqnarray}
Here, $N = \frac{4s^2}{\Delta a}$ for the 4 $X$/$Y$ faces and
$\bar{l^2}$ is a mean square armlength, with $\bar{l^2} = \frac{2}{s}
\int_0^{s/2} x^2 \, dx = \frac{s^2}{12}$ for $\Delta a \ll s^2$. This is
approximate in the case of only a few uncorrelated domains per TM face,
such as for $\Delta a$ the size of an $X$ electrode ($\approx
500$~mm$^2$). 

Table \ref{supp_table} summarizes the statistical variation for
$\Delta_{\phi}$ compared to that of $\Delta_{x}$ for different
assumptions of the relevant minimum area $\Delta a$ beyond which 
the average surface potentials become uncorrelated. 
In the limit of
small correlation-length domains ($\Delta a \ll s^2$), the variance
$\langle \Delta_{\phi} ^2 \rangle \approx 3 \langle \Delta_x ^2
\rangle$, with a factor 2 from doubling, from 2 to 4, the faces with
gap-dependent contributions to $\Delta_{\phi}$ and a factor
$\frac{s^2}{12 R_x^2} \approx 1.5$ for the average square armlength
compared to that of a single $X$ electrode. 

\begin{table}[b]
\centering
\begin{ruledtabular}
\begin{tabular}{ | c | c | c | c |}

\, & $\Delta a \ll s^2$  & $ \Delta a \approx s^2$  & equipotential
\\
\, & (or $\Delta a \approx A_x $)  & \,  & conductors
\\  
\colrule
\, & \, & \, & \,  \\
$\frac{\langle \Delta_{\phi}^2 \rangle}{\langle \Delta_x^2 \rangle}$

& $\frac{s^2}{6 R_x^2} \approx 3$

& -- -- -- 

& 1

\\
\, & \, & \, & \,  \\ 
\colrule 
\, & \, & \, & \,  \\
$\frac{\langle V_{UW}^2 \rangle}{\langle \Delta_x^2 \rangle}$
& $ \frac{2 A_x^2}{3 A s^2} \approx 0.04 $

& $\frac{A_x^2}{2 s^4} \approx 0.03 $

& $ \approx 0.03 $
\\
\, & \, & \, & \,  \\ 
\end{tabular}
\end{ruledtabular}
\caption{\label{supp_table}Values for scaling potential differences in 
torque measurements of $\Delta_{\phi}$ and $V_{UW}$ 
to the $\Delta_x$ relevant to the interaction with TM
charge in LISA. We consider cases in which the minimum 
surface area of the underlying characteristic
domains, beyond which potentials become uncorrelated, are (1) smaller
than the TM cube sidelength $s$, (2) roughly equal to the TM dimension
$s$, and (3) coincide with the individual conducting surface
boundaries -- TM and electrodes for LISA and the individual
plates in the configuration of Ref. \cite{pollack_prl} -- so that each has a
single uniform potential. As discussed in the text, $\Delta_{\phi}$ is 
not a statistical indicator of $\Delta_x$ in the case ($\Delta a \, \approx s^2$)
that each TM and sensor face is a unique equipotential.  
Cases 2 and 3 coincide for $V_{UW}$, in which the only two relevant 
conductors have $A \approx s^2$.}
\end{table}

If, instead, all distinct conducting surfaces -- for LISA, the TM, the
individual sensor electrodes, and the rest of the electrode housing --
were individually equipotential, then $\Delta_x$ and $\Delta_{\phi}$ are
both determined only by the stray potential values on the 4 $X$
electrodes, as these are the only full conductors with a non-vanishing
$\pder{C}{x}$ and $\pder{C}{\phi}$. $\Delta_x$ and $\Delta_{\phi}$ are
thus different combinations of these 4 potentials in this case, as in the 
simplified analysis of Ref. \cite{DC_bias_asr}, and
their expected statistical variances are equal. 

In the case that each entire sensor and TM face is a distinct
equipotential, then $\Delta_{\phi} = 0$, as the contribution of one half
of a sensor (or TM) face cancels that of the other half, 
with $\pder{C}{\phi}$ changing sign. $\Delta_{x}$
could still be non-zero in this case, and thus measurement of
$\Delta_{\phi}$ would no longer be a good indicator of the statistics of
$\Delta_x$. This case is, however, considered highly unlikely in the
case of the LISA prototype sensor under study, where the individual
faces of the sensor, including electrodes and surrounding guard ring
surfaces, are composed of physically separate conductors, whose gold
coatings are connected eletrically only through the attached circuitry.
As such, the distribution of $\Delta_{\phi}$, and its fluctuations, are
taken as a statistical indicator for $\Delta_x$ that, considering Table
\ref{supp_table}, is slightly pessimistic over a range of assumptions
for the underlying potential distribution. 

It is interesting to scale the measured values for the noise
$S_{\Delta_{\phi}}$ to the corresponding noise in the average potential
on a conductor of given size, for instance that of a single LISA $X$
electrode. Substituting $\Delta a = A_x$ in \eqr{s_Delta_phi_model} and
approximating $\bar{l^2} \approx \frac{s^2}{12}$, we find $\langle
\Delta_{\phi} ^2 \rangle \approx V_0^2 \frac{s^4}{3 A_x R_x^2} \approx
25 V_0^2$ (corresponding to a relevant surface area $16 \times A_x$ on
the 4 $X$ and $Y$ faces and the factor 1.5 armlength correction mentioned
above). As such, in the limit that the potential fluctuations are
correlated only on a scale smaller than $A_x \approx 500$~mm$^2$, the
measured noise in $\Delta_{\phi}$ is roughly 5 times larger, in linear
spectral density, than the noise in the average potential difference 
between opposing 500 mm$^2$ surfaces. 

In the geometry studied in Ref. \cite{pollack_prl}, shown at right in
\figr{supp_sensor_fig}, a modulated voltage is applied between
two parallel plates, with relevant overlap width $\frac{L}{2}$ and
height $h$ (and thus area $A = \frac {h L}{2}$). 
The measured torque is converted into an equivalent
potential difference between the two plates by dividing by the total
capacitive derivative $\pder{C}{\phi}$. Given this normalization and the
electrostatic model used in our text, the measured potential difference
in these measurements, $V_{UW}$, is given by
\begin{equation}
V_{UW} = \frac{1}{\left| \pder{C}{\phi} \right|}
\sum_{i \left(1 \right),j \left( 2 \right) }
{\! \! \! \pder{C_{ij}}{\phi} \left( \delta V_j  - \delta V_i \right) }
\label{s_UW}
\end{equation}

Following the same analysis applied for $\Delta_x$ and $\Delta_{\phi}$,
with  $\pder{C}{\phi} \approx \frac{A \epsilon_0 \frac{L}{4}}{d^2}$,
\begin{eqnarray}
V_{UW} & \approx & \frac{\Delta a}{A \frac{L}{4}} \sum_{m}{l_m V_m}
\nonumber \\
\langle V_{UW}^2 \rangle 
& \approx & 
\Delta a \, V_0^2 \frac{16 \bar{l^2}}{A L^2} 
\label{s_UW_model}
\end{eqnarray}

The surface area $A \approx 2180$~mm$^2$ used in the measurement -- 
and in the normalization of $V_{UW}$ -- is roughly 
that of a LISA TM face, $s^2$, and four times that of a LISA sensor $X$
electrode. In rough terms, the mean square variance in $V_{UW}$ will be
smaller than that of $\Delta_x$ by a factor 16 -- due to the factor 4
in normalization area, $\frac{A}{A_x}$ (see Eqns. \ref{s_Delta_x_I},
\ref{s_ave_Delta_x} and \ref{s_UW},\ref{s_UW_model}) -- and by an
additional statistical factor 2, for the ratio of relevant surface area, 
$\frac{2 s^2}{A}$, which means half the number of domains in the
UW geometry, regardless of their size. Including a small armlength
correction factor weighing domains farther from the torque axis, which
varies from 1 for $\Delta a \, \approx s^2$ to $\frac{4}{3}$ for 
$\Delta a \, \ll s^2$, the standard deviation in $V_{UW}^2$ is 25-35 
times smaller than that for $\Delta_x^2$ (see Table \ref{supp_table}). 
As such, to deduce an expectation value for the noise in
$\Delta_x$ based on potential fluctuation measurements in the UW
geometry, we must scale the linear noise density by a factor 5-6.

% \bibliography{charge_DC_bib}

\end{document}